# New Type Phase Transition of Li$_2$RuO$_3$ with Honeycomb Structure

Yoko Miura, Yukio Yasui, Masatoshi Sato*, Naoki Igawa[1] and Kazuhisa Kakurai[1]

*Department of Physics, Nagoya University, Furo-cho, Chikusa-ku, Nagoya 464-8602*
[1]*Quantum Beam Science Directorate, Japan Atomic Energy Agency,*
*Tokai-mura, Naka-gun, Ibaraki 319-1195.*

**Abstract**

A new-type structural transition has been found in Li$_2$RuO$_3$ with honeycomb lattice of edge-sharing RuO$_6$-octahedra. With decreasing temperature $T$, the electrical resistivity exhibits an anomalous increase at $T=T_c\sim 540$ K, suggesting the (metal to insulator)-like transition and the magnetic susceptibility also shows a sharp decrease. Detailed structure analyses have revealed that the high temperature space group $C2/m$ changes to $P2_1/m$ at $T_c$. The most striking fact is that a significant reduction of the bond lengths is found between two of six Ru-Ru pairs of the hexagon in the low temperature phase, indicating a new type phase transition by the mechanism of the formation of molecular orbits of these Ru-Ru pairs.

KEYWORDS: Li$_2$RuO$_3$, honeycomb structure, structural transition

*Corresponding author: e43247a@nucc.cc.nagoya-u.ac.jp

Compounds with the honeycomb lattice often present interesting behavior originating from their characteristic structures. For example, in the course of the studies on the physical properties of localized spin systems of A$_3$T$_2$SbO$_6$ (A=Na, Li; T=Cu, Ni, Co) and Na$_2$T$_2$TeO$_6$ on the (distorted) honeycomb lattice, spin gap behaviors have been found for T=Cu,[1-3] while the magnetic transitions to the spin-ordered state have been observed for T=Co and Ni.[4]

As one of possible examples of conductive electrons on honeycomb lattice, we have investigated physical properties of Li$_2$RuO$_3$. It has the layers of the honeycomb lattice of edge-sharing RuO$_6$ octahedra with a LiO$_6$ octahedron at the center of each hexagon of RuO$_6$ (Fig. 1). The Ru valence is +4 and the four electrons exist in the $4d\ t_{2g}$ orbits. For this system, we have found a phase transition at temperature $T=T_c\sim 540$ K, where the crystal symmetry changes from a monoclinic (space group $C2/m$) to another monoclinic (space group $P2_1/m$) one with decreasing $T$. As described in detail later, the transition is associated with the molecular orbit formation of Ru$^{4+}$-Ru$^{4+}$ ions of the edge-sharing RuO$_6$ pair, presenting a new mechanism of structural transitions.

Polycrystalline samples of Li$_2$RuO$_3$ were prepared by sintering pellets of mixtures of RuO$_2$ and Li$_2$CO$_3$ with proper molar ratio at 1000 °C for 24 h in air.[5,6] The powder neutron diffraction patterns of these samples indicate that a small amount of RuO$_2$ (molar fraction of ~1.20 %) exists. There also exists an impurity peak of the unidentified phase, which has the integrated intensity of ~4.5 % of the maximum integrated intensity of the main phase as shown later. The magnetic susceptibilities $\chi$ were measured using a Quantum Design SQUID magnetometer under a magnetic field $H=1$ T in the temperature range of 2-700 K. The electrical resistivities $\rho$ were measured by the standard four-terminal method using an ac-resistance bridge from 4.6 K to 695 K. The specific heats $C$ were measured by a thermal relaxation method in the temperature range of 5-60 K using a Physical Property Measurement System (PPMS, Quantum Design). Powder X-ray diffraction measurements were carried out with Cu K$\alpha$ radiation. Powder neutron diffraction measurements were carried out at room temperature (RT) and 600 K using the high-resolution powder diffractometer (HRPD) installed at JRR-3 of JAEA in Tokai. The 331 reflection of Ge monochromator was used. The horizontal collimations were open (35′)-20′-6′ and the neutron wavelength λ was ~1.824 Å. The diffraction intensities were measured in the 2θ-range from 2.5 to 165° at a step of 0.05°. Rietveld analyses were carried out for these data using RIETAN-2000.[7]

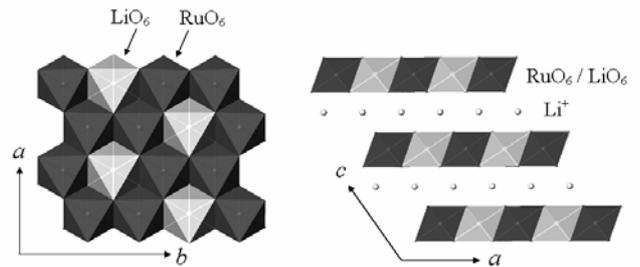

Fig. 1  Schematic figures of Li$_2$RuO$_3$ viewed from the direction perpendicular to the *ab*-plane (left) and from the *b*-direction (right). At the corners of the octahedra, O atoms exist, and a Ru or Li atom is within each octahedron. Li atom layers are between the Ru$_2$LiO$_6$ layers.

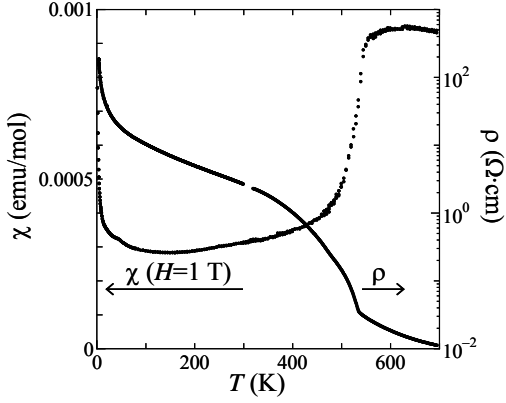

Fig. 2  Magnetic susceptibility χ measured under the magnetic field $H=1$ T and the electrical resistivity ρ of $Li_2RuO_3$ are shown against $T$.

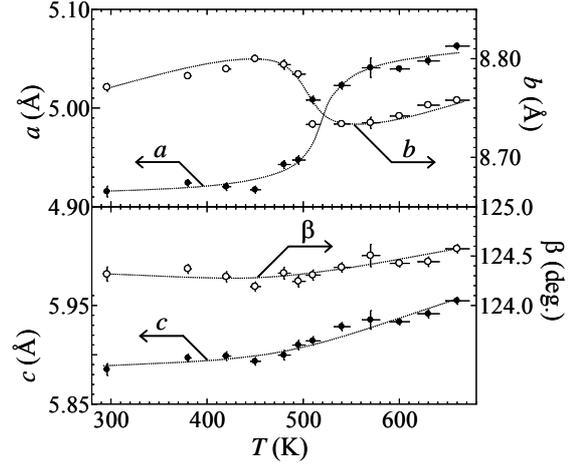

Fig. 3  Lattice parameters obtained by powder X-ray diffraction are shown against $T$, where the dotted lines are guides for the eyes.

The results of the measurements of χ and ρ of the polycrystalline samples of $Li_2RuO_3$ are shown in Fig. 2 against $T$. A steep decrease of χ has been found at $T_c$~540 K with decreasing $T$. The increase of χ found with decreasing $T$ in the low $T$ region is possibly due to the existence of impurities or other kinds of lattice imperfections. The resistivity ρ exhibits a steep increase at $T_c$ with decreasing $T$. We could not check if the system exhibits the metallic $T$ dependence in the $T$ region far above $T_c$ where no critical fluctuation of the transition exists, because the system is not stable above the highest temperature studied here.

Measuring the specific heat of $Li_2RuO_3$ (the data not shown), we have estimated the electronic specific heat coefficient γ to be ~1.40 (mJ/mol·K$^2$). Because the value is much smaller than the value of 22 mJ/mol·K$^2$ obtained by assuming that the χ value observed at low temperatures is due to the Pauli spin susceptibility, almost whole part of the low $T$ value of χ is considered to be the contribution from the Van Vleck paramagnetism. We do not know if the nonzero value of γ is due to the existence of the small and intrinsic Fermi surface or it is just due to the small amount of metallic parts existing in the samples.

The powder X-ray measurements were carried out and the obtained lattice parameters are shown in Fig. 3 against $T$. The lattice parameter $a$ ($b$) decreases (increases) rapidly at $T_c$ with decreasing $T$. The value of $b/a$ is ~√3 above $T_c$, indicating that the honeycomb structure changes from a nearly ideal form to a distorted one as the result of the transition.

Neutron Rietveld analyses have been carried out at 600 K and RT (see Figs 4 and 5). Although the space group at RT was reported previously to be $C2/c$,[5,6] it cannot explain the superlattice peaks indicated by the black arrows in Fig. 5. To reproduce these reflections, we have to adopt the space group $P2_1/m$, for which the conventional unit cell with a half volume of that for $C2/c$ can be used. (The cell for $P2_1/m$ has the single Ru-honeycomb layer, while that for $C2/c$ has two Ru-honeycomb layers.) The result of the Rietveld fitting by $P2_1/m$ is rather well. As the possible space group at 600 K, we have adopted $C2/m$, the minimal non-isomorphic supergroup of $P2_1/m$, because for this space group, we can take the unit cell with the single $RuO_6$-honeycomb layer and because it allows the second order transition at $T_c$. (From the experimental data, it is not easy to definitely distinguish if it is the second order one or the first order one.)

In the fitting, we have obtained satisfactory results (Figs. 4 and 5). As stated above, the superlattice peaks observed at RT(<$T_c$) are indicated by the black arrows. (The peaks from $RuO_2$ and the unidentified phase are indicated by the open and gray arrows, respectively.) The obtained $R$ factors are as follows. At RT, $R_{wp}$=4.17, $R_e$=3.07 and $S$=1.36 for $P2_1/m$. At 600K, $R_{wp}$=4.32, $R_e$=3.07 and $S$=1.41 for $C2/m$. The lattice constants and atomic coordinates at 600 K for $C2/m$ are as follows. $a$=5.0466(3) Å; $b$=8.7649(2) Å; $c$=5.9417(3) Å; β=124.495(4)°; Ru (4g) $y$=0.3308(3); Li1 (2a); Li2 (4h) $y$=0.3425(9); Li3 (2c); O1 (8j) $x$=-0.0162(7), $y$=0.1701(2), $z$=0.2324(2); O2 (4i) $x$=0.4991(8), $z$=0.2325(4). Those at RT for $P2_1/m$ are $a$=4.9210(2) Å; $b$=8.7829(2) Å; $c$=5.8941(2) Å; β=124.342(2)°; Ru (4f) $x$=0.2737(6), $y$=0.0766(2), $z$=-0.0063(6); Li1 (2e) $x$=0.706(3), $z$=-0.068(2); Li2 (4f) $x$=0.253(3), $y$=0.0991(7), $z$=0.493(3); Li3 (2e) $x$=0.772(4), $z$=0.513(5); O1 (4f) $x$=0.7553(7), $y$=0.0805(6), $z$=0.2489(6); O2 (4f) $x$=0.7757(6), $y$=0.0819(6), $z$=0.7685(6); O3 (2e) $x$=0.255(1), $z$=0.2144(9); O4 (2e) $x$=0.267(1), $z$=0.761(1). We think that the space group is $C2/m$ above $T_c$ and the second order transition takes place at $T_c$. If the transition is the first order one, the space group at 600 K might be $C2/c$. (The Rietveld fitting is equally well for $C2/c$ at 600 K. The obtained $R$ factors are $R_{wp}=4.27$, $R_e=3.07$ and $S=1.39$ for $C2/c$.)

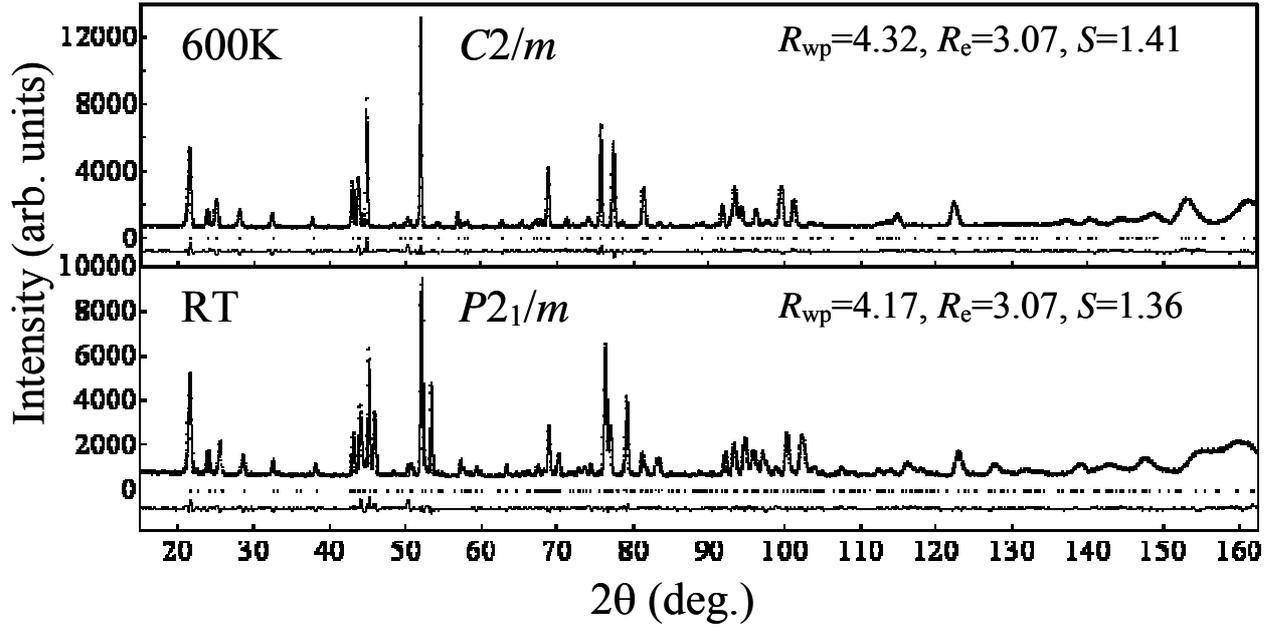

Fig. 4 Neutron powder diffraction patterns of $Li_2RuO_3$ at 600 K (upper panel) and RT (lower panel). The data are shown by the crosses and the calculated curves obtained by the Rietveld fittings with monoclinic ($C2/m$ at 600K and $P2_1/m$ at RT) structures are shown by the solid curves. The Bragg positions are indicated by the vertical lines and the differences between the observed and calculated data are shown at the lowest part of each panel.

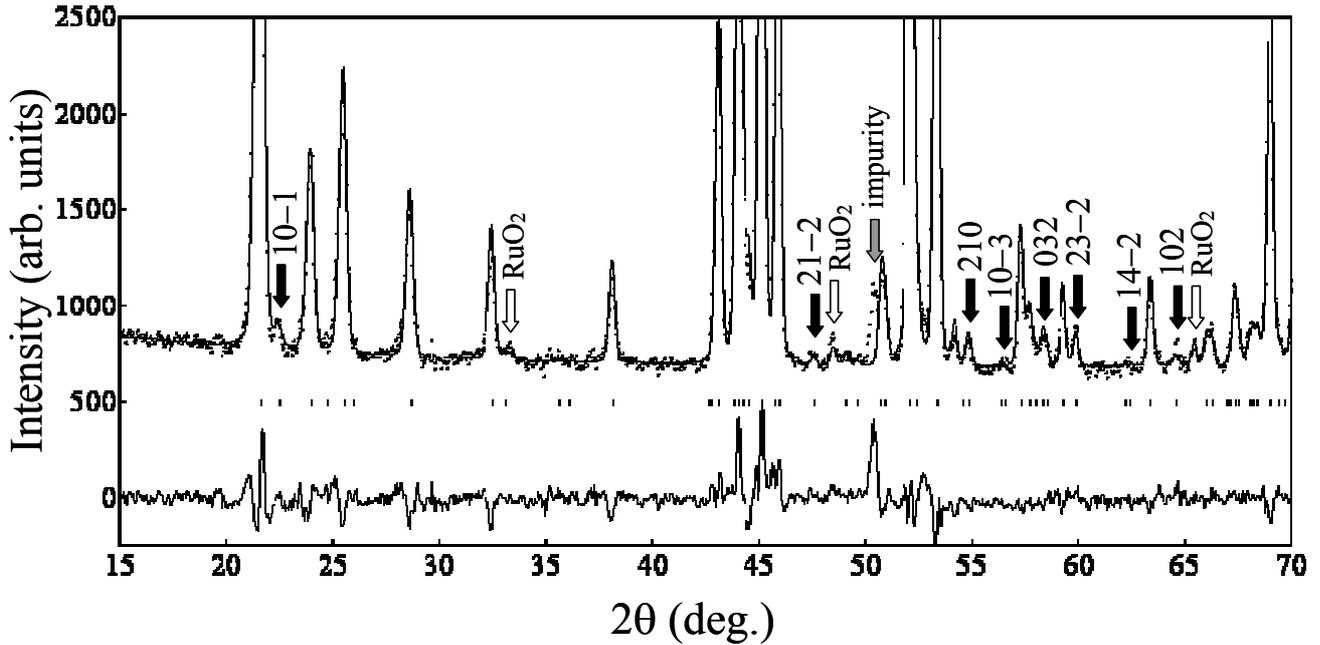

Fig. 5 Neutron powder diffraction pattern of $Li_2RuO_3$ at RT is shown with the enlarged scales. The data are shown by the crosses and the calculated curves obtained by the Rietveld fitting with monoclinic ($P2_1/m$) structures are shown by the solid curve. Black arrows indicate the superlattice reflections observed at RT which have relatively large intensities. The open and gray arrows indicate peaks from the $RuO_2$ and an unidentified impurity phase, respectively. Bragg positions are indicated by the vertical lines and the differences between the observed and calculated data are shown at the lowest part.

Figure 6 shows Ru-honeycomb skeletons at 600 K and RT. At 600 K, Ru atoms form almost ideal honeycomb structure with the Ru-Ru distances $a_1=2.966$ Å and $a_2=2.894$ Å, where the difference between $a_1$ and $a_2$ is just ~2.5 % of these distances. At RT, the hexagons are distorted very significantly: $a_1=3.045$ Å, $a_2=3.049$ Å and $a_3=2.568$ Å, where $(a_2-a_3)/a_3 \sim 19$ %. The value of $a_3$ is reduced by an amount as large as 0.326 Å.

Why do such drastic changes take place? To answer the question, we consider the coupling of the lattice distortion with the formation of the molecular orbits of the $t_{2g}$ electrons of $Ru^{4+}$-$Ru^{4+}$ of edge-sharing $RuO_6$ octahedra. The σ-, π- and δ-molecular orbits (bonding orbits) schematically shown in the lower part of Fig. 7, are, we think, formed by the three pairs of the $t_{2g}$ orbits.

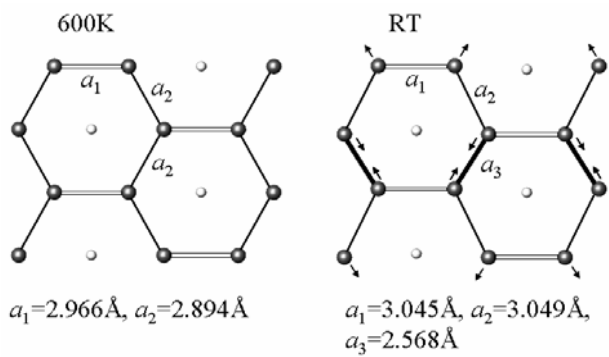

Fig. 6 The Ru honeycomb skeletons at 600 K (left) and RT (right) and the Ru-Ru atomic distances $a_1$, $a_2$ and $a_3$ are shown in each figure. At RT, $a_3$ is smaller than that at 600 K by ~13 %.

The corresponding antibonding states are $\sigma^*$-, $\pi^*$- and $\delta^*$-orbits, respectively. The energy levels of these orbits are shown in the upper part of the figure and we find that the nonmagnetic electron configuration is realized as shown in the figure. The reduction of the atomic distance between the $Ru^{4+}$-$Ru^{4+}$ pair is, we think, induced by this change of the electronic state. The increase of $\rho$ can be naturally understood by this mechanism. Even in the case where the space group above 540 K is $C2/c$, the similar mechanism can be considered to be relevant.

Although transitions accompanied with the significant anomalies of $\chi$ and $\rho$ similar to those of the present system, have been reported for $La_4Ru_2O_{10}$[8-10] with corner-sharing $RuO_6$ octahedra and $Tl_2Ru_2O_7$[11] with pyrochlore structure, for example, their microscopic mechanisms are different from the presently proposed one. Only for $La_4Ru_6O_{19}$ and $Bi_3Ru_3O_{11}$ with the edge-sharing $RuO_6$ octahedra, the similar mechanism has been discussed to understand their $T$ dependent magnetic susceptibilities.[12-15] However, these systems do not exhibit a phase transition. The $T$ dependent susceptibility due to the similar electronic configuration has also been reported for various complexes with metal ion pairs.[16,17] Of course, these complexes do not exhibit phase transitions. $Li_2RuO_3$ is considered to present, as far as we know, a new type of phase transitions associated with the structural distortion induced by the molecular orbit formation of the $d$ electrons. In this sense, it is interesting to study the band structure of this system at both temperatures above and below $T_c$.

**Acknowledgement**

This work is supported by Grants-in-Aid for Scientific Research from the Japan Society for the Promotion of Science (JSPS) and by Grants-in-Aid on Priority Areas from the Ministry of Education, Culture, Sports, Science and Technology.

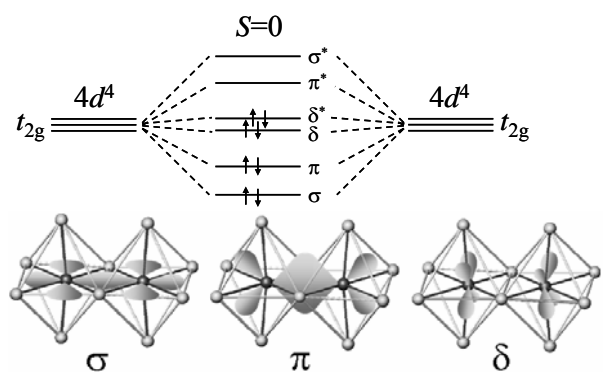

Fig. 7 (Upper panel) The energy levels of molecular orbits of $Ru^{4+}$- $Ru^{4+}$ pairs. The electrons in these levels are shown by the arrows presenting the spin directions. (Lower panel) The schematic figures of the wave functions of the σ-, π- and δ-molecular orbits (bonding orbits) formed by the three pairs of the $t_{2g}$ orbits. The black and gray circles are Ru and O atoms, respectively.